\documentclass[epjCONF]{svjour}
\usepackage{graphics}
\usepackage[varg]{txfonts} 
\usepackage[latin1]{inputenc}
\session-title{The Space Photometry Revolution, CoRoT3-KASC7 joint meeting}
\def\nat{Nature\ }
\def\prl{Phys.\ Rev.\ Lett.\ }
\def\aap{Astron.\ Astrophys.\ }

\def\apj{Astrophys.\ J.\ }
\def\apjl{Astrophys.\ J.\ Lett.\ }

\def\mnras{Mon.\ Not.\ Roy.\ Astron.\ Soc.\ }

\def\prd{Phys.\ Rev.\ D\ }

\begin{document}
\title{Dark matter searches with asteroseismology}
\author{Jordi Casanellas\inst{1}\fnmsep\thanks{\email{jordi.casanellas@aei.mpg.de}}}
\institute{Max Planck Institute for Gravitational Physics, Albert Einstein Institute, Am M\"uhlenberg 1, 14476, Golm, Germany}
\abstract{
High-precision asteroseismic data provides an excellent opportunity to test theories of stellar evolution and new physics, such as the properties of the dark matter (DM) of the Universe. Here we will show that some models of DM lead to changes in the classical scenario of stellar evolution. The accumulation of DM in the core of low-mass stars reduces their central temperatures and inhibits the formation of small convective cores in 1.1-1.3 Ms stars. We review the asteroseismic constraints that have been set to the characteristics of the DM particles, obtained comparing the oscillations of the star $\alpha$~Cen~B with modified stellar models. To conclude, we discuss the prospects to use CoRoT and Kepler data on main-sequence stars and red giants to further constrain the nature of DM.} 
\maketitle
\section{Introduction}
\label{intro}
The discovery of the nature of Dark Matter (DM) remains as one of the main challenges in our understanding of the Universe, and thus is a key goal both for astrophysicists, cosmologists, and particle physicists. Current experimental efforts, including direct and indirect detection techniques and DM searches at the LHC, have already reached the sensitivity to detect particular models of WIMPs (for Weakly Interacting Massive Particles), with no positive success at the moment, although several hints of detection have been claimed. In fact, one of the reasons for WIMPs to be very popular DM candidates is that their existence could be detected in ongoing or upcoming experiments~\cite{Bertone:2010at}.\\

Stellar physics and observations have provided fruitful insights in particle physics and cosmology~\cite{Perlmutter,Casanellas2012}, triggering advances in the understanding of neutrino interactions and oscillations~\cite{BahcallBethe1993}, and setting bounds on new particles like WIMPs~\cite{2013ApJ...765L..21C}, axions~\cite{2008ApJ...682L.109I} and other light bosons~\cite{2013MNRAS.432.3332V}. In this context, the unprecedented abundance of high-quality data from asteroseismic missions like CoRoT~\cite{Michel:2008im} and Kepler~\cite{Bruntt:2012hs} opens up the possibility of further use the stars as laboratories for fundamental physics~\cite{1996slfp.book.....R}.\\

\section{Impact of dark matter on stars}
Stars are formed and evolve within clouds of DM particles. If these particles are WIMPs, they would efficiently accumulate inside stars and could potentially influence their internal structure~\cite{1985ApJ...294..663S,art-Gould1987}. This hypothetical impact depends on the unknown characteristics of the DM particles like their mass, $m_{\chi}$, and their scattering cross-section with protons, $\sigma_{\chi,SD}$~\cite{Lopes:2011rx}.\\

On the assumption that the DM sector has an asymmetry in the abundances of particles and antiparticles like that observed in the normal matter~\cite{Kaplan:2009ag} (or, equivalently for our purposes, a very small self-annihilation cross section, $\langle \sigma_a v \rangle \lesssim 10^{-33}\;$cm$^3\;$s$^{-1}$) the impact on stellar evolution can be noticeable. Although the density of DM in our region of the Galaxy is estimated to be very low, $\rho_\chi \approx 0.4\;$GeV~cm$^{-3}$ \cite{Garbari:2012ff}, it suffices to change the internal properties of stars similar to the Sun~\cite{2013ApJ...765L..21C}. In environments with greater DM densities, such as the center of the Milky Way and inside dwarf galaxies and globular clusters, the impact on stars could be much larger~\cite{Casanellas:2009dp,Casanellas:2010he,Casanellas:2011qh,Iocco:2012wk}, in particular in the case of compact stars~\cite{Kouvaris:2011fi,Travis2014,FullerOtt2015,Bramante2015}.\\

\begin{figure}
\centering
\includegraphics{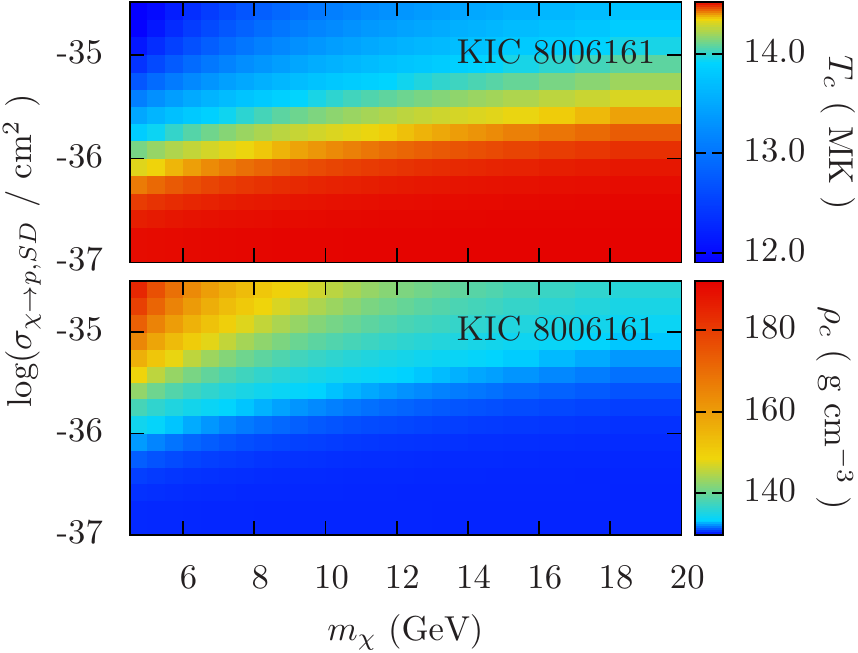}
\caption{Central temperatures (top) and densities (bottom) of the DM-modified stellar models that reproduce the observed properties of the star KIC 8006161, for different masses $m_{\chi}$ and spin-dependent scattering cross sections on protons $\sigma_{\chi,SD}$ of the DM particles. See Ref.~\cite{2013ApJ...765L..21C} for further details.}
\label{fig:TcRhoc}       
\end{figure}

The DM particles trapped in the stellar cores transport energy by conduction~\cite{1986ApJ...306..703G}, removing energy from the center of the star and bringing it to outer layers at a rate, 
\begin{eqnarray}
L_{\chi,trans}(r)= 4 \pi r^2 n_{\chi}(r) l_{\chi}(r) \kappa(r) \left(\frac{k_B T(r)}{m_{\chi}}\right)^{1/2} k_B \frac{dT}{dr} f(K) h(r),
\end{eqnarray}
which depends, among other factors, on the WIMPs number density distribution, $n_{\chi}(r)$, and their mean free path inside the star, $l_{\chi}(r)=\left(\sum \sigma_{\chi,i} n_i(r)\right)^{-1}$ (see Refs.~\cite{1990ApJ...352..654G,Scott:2008ns} for more details). As a result of this new cooling mechanism, low-mass stars have their central temperatures reduced and their central densities increased when compared to the classical theory of stellar evolution (see Ref.~\cite{2013ApJ...765L..21C} and Fig.~\ref{fig:TcRhoc}). This phenomena has been studied in the case of the Sun, where the observations of solar neutrinos and helioseismology have been used to constrain the properties of DM~\cite{LopesSilk2010Sc,2012ApJ...746L..12T,2012ApJ...752..129L}.\\

\section{Asteroseismic diagnostic of Dark Matter}
Compared to solar observations, asteroseismology provides the opportunity to study stars with different masses and at different evolutionary stages. Regarding DM searches, one of the main advantages is the possibility to analyse stars with masses lower than the Sun, which are more strongly influenced by DM conduction than our star. On the other hand, one should be aware that the precision on the determination of stellar characteristics, like the mass and metallicity, is crucial to disentangle the impact of DM from other standard mechanisms.\\

\begin{figure}
\centering
\includegraphics{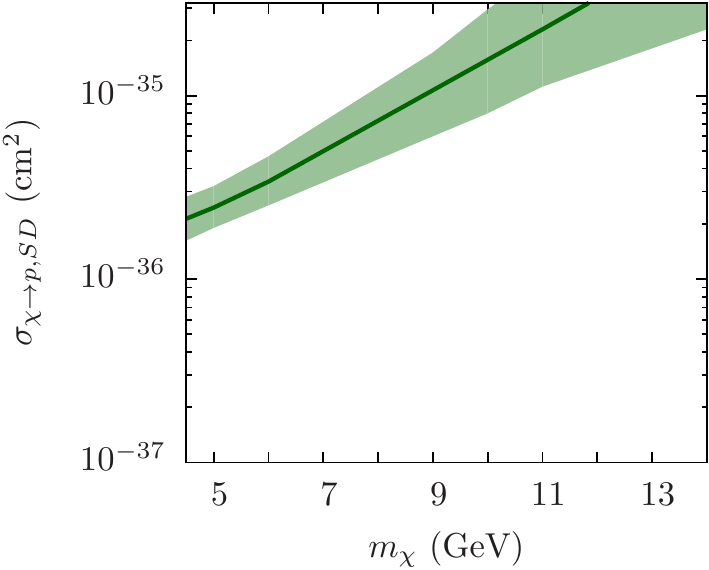}
\caption{Upper limits for the DM-proton spin-dependent scattering cross section, $\sigma_{\chi,SD}$, as a function of the DM mass, $m_{\chi}$, from an asteroseismic analysis of the star $\alpha$~Cen~B. The existence of asymmetric DM particles with properties above the blue line can be ruled out because they produce a strong impact on the core of the star, leading to a mean small frequency separation, $\delta \nu_{02}$, more than 2~$\sigma$ away from the observations. The filled region shows the uncertainty in the modelling when the observational errors are taken into account. See Ref.~\cite{2013ApJ...765L..21C} for further details.}
\label{fig:mxsigx}       
\end{figure}

In Ref.~\cite{2013ApJ...765L..21C} we studied $\alpha$~Cen~B, a 0.9~M$_{\odot}$ star which characteristics have been accurately estimated because it belongs to a binary system. We found that some models of DM lead to changes in the central properties of the star that are in conflict with its acoustic oscillations. In particular, the small frequency separation between modes of low degree,
\begin{equation}
\delta \nu_{02}=\nu_{n,0}-\nu_{n-1,2} \;,
\end{equation}
was found to be in conflict with the observations for the modified stellar models with DM particles with small $m_{\chi}$ and large $\sigma_{\chi,SD}$ (see Fig.~\ref{fig:mxsigx}).\\

\begin{figure}
\centering
\resizebox{0.95\columnwidth}{!}{%
\includegraphics{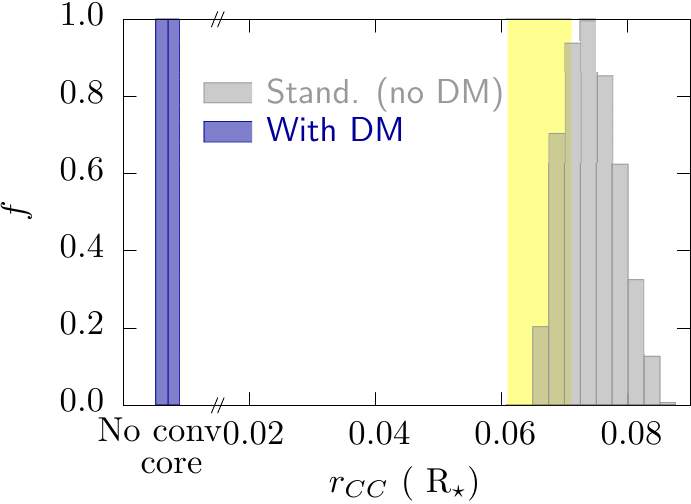}
\includegraphics{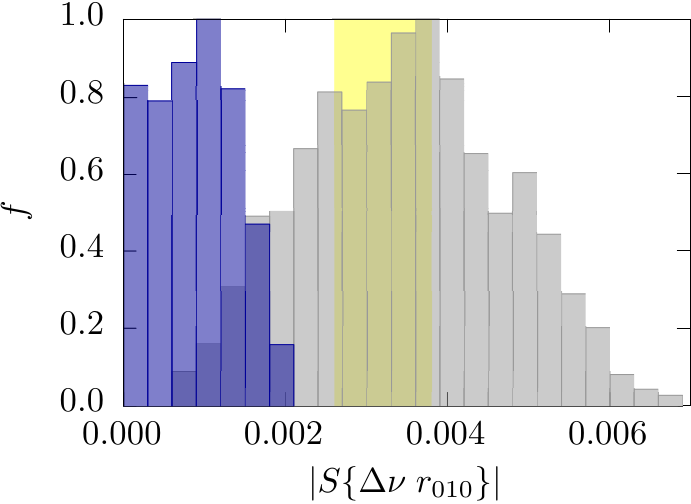} }
\caption{Left: Size of the convective core of star KIC 12009504 obtained changing its mass, metallicity, and mixing-length parameter for the standard scenario without DM (grey) and including the impact of asymmetric DM particles with the properties to explain recent claims in direct detection experiments (blue). Right: Seismic parameter $|S\{ \Delta\nu \; r_{010}\}|$ for the same models. The yellow areas show the $r_{cc}$ derived by Ref.~\cite{SilvaAgDushera} and $|S\{ \Delta\nu \; r_{010}\}|$ calculated from the observed oscillation frequencies. See Ref.~\cite{CasanellasBrandao} for further details.}
\label{fig:rccsl}       
\end{figure}

Another strong impact of asymmetric DM on main-sequence stars occurs for stars with masses around 1.2~M$_{\odot}$. These stars can have their small convective cores suppressed due to the efficiency of the extra energy transport by DM conduction. In a recent work~\cite{CasanellasBrandao}, it was shown that the star KIC 12009504 (also called Dushera) would not have a convective core if certain DM particles exist and, consequently, asteroseismic parameters like frequency ratios~\cite{art-RoxburghVorontsov2003A&A} or parameters such as $S\{ \Delta\nu \; r_{010}\}$~\cite{BrandaoCunhaJCD} can be used to put limits to the properties of DM (see Fig.~\ref{fig:rccsl}).\\

Further constraints on the nature of DM could be set if solar-like oscillations are identified in stars with masses significantly lower than the Sun, such as M dwarf stars. Unfortunately, their faintness and the expected low-amplitude of their oscillations has prevented the detection of pulsations in these low-mass stars~\cite{Rodriguez-Mdwarfs}. On the other hand, the recent measurement of gravity modes in the oscillations of Red Giants could also be used to investigate DM because it allows a precise insight in the core of these stars, where DM may be having an impact.\\

\begin{acknowledgement}
The author acknowledges the support from the Alexander von Humboldt Foundation.
\end{acknowledgement}


\begin{thebibliography}{}
\bibitem{Bertone:2010at}
  Bertone, G., Nature 468, (2010) 389-393

\bibitem{Perlmutter}
  Perlmutter, S., Aldering, G., Della Valle, M. \textit{et al.}, \nat 391, (1998) 51-54

\bibitem{Casanellas2012}
  Casanellas, J., Pani, P., Lopes, I. and Cardoso, V., \apj 745, (2012) 15
 
\bibitem{BahcallBethe1993}
  Bahcall, J. and Bethe, H., \prd 47, (1993) 1298-1301
 
\bibitem{2013ApJ...765L..21C}
  Casanellas, J. and Lopes, I., \apjl 765, (2013) L21
 
\bibitem{2008ApJ...682L.109I}
  Isern, J., Garc{\'{\i}}a-Berro, E., Torres, S. and Catal{\'a}n, S., \apjl 682, (2008) L109-L112

\bibitem{2013MNRAS.432.3332V}
  Vincent, A.~C., Scott, P. and Trampedach, R., \mnras 432, (2013) 3332-3339
  
\bibitem{Michel:2008im}
 Michel, E., Baglin, A., Auvergne, M., Catala, C. and Samadi, R., Science 322, (2008) 558-560
 
\bibitem{Bruntt:2012hs}
  Bruntt, H., Basu, S., Smalley, B. \textit{et al.}, \mnras 423, (2012) 122-131
  
\bibitem{1996slfp.book.....R}
  Raffelt, G.~G., \textit{Stars as laboratories for fundamental physics : the astrophysics of neutrinos, axions, and other weakly interacting particles} (University of Chicago Press, Chicago 1996)

\bibitem{1985ApJ...294..663S}
  Spergel, D. and Press, W., \apj 294, (1985) 663-673

\bibitem{art-Gould1987}
  Gould, A., \apj 321, (1987) 571--585
  
\bibitem{Lopes:2011rx}
  Lopes, I., Casanellas, J. and Eugenio, D., \prd 83, (2011) 063521

\bibitem{Kaplan:2009ag}
  Kaplan, D., Luty, M. and Zurek, K., \prd 79, (2009) 115016
  
\bibitem{Garbari:2012ff}
  Garbari, S., Liu, C., Read, J. and Lake, G., \mnras 425, (2012) 1445-1458

\bibitem{Casanellas:2009dp}
  Casanellas, J. and Lopes I., \apj 705, (2009) 135-143

\bibitem{Casanellas:2010he}
  Casanellas, J. and Lopes I., \mnras 410, (2011) 535-540
  
\bibitem{Casanellas:2011qh}
  Casanellas, J. and Lopes I., \apjl 733, (2011) L51

\bibitem{Iocco:2012wk}
  Iocco, F., Taoso, M., Leclercq, F. and Meynet, G., \prl 108, (2012) 061301
  
\bibitem{Kouvaris:2011fi}
  Kouvaris, C. and Tinyakov, P., \prl 107, (2011) 091301

\bibitem{Travis2014}
  Hurst, T., Zentner, A., Natarajan, A. and Badenes, C., (2014) arXiv:1410.3925
  
\bibitem{FullerOtt2015}
  Fuller, J. and Ott, C., \mnras 450, (2015) L71

\bibitem{Bramante2015}
  Bramante, J, (2015) arXiv:1505.07464 
  
\bibitem{1986ApJ...306..703G}
  Gilliland, R., Faulkner, J., Press, W. and Spergel, D., \apj 306, (1986) 703-709

\bibitem{1990ApJ...352..654G}
  Gould, A. and Raffelt, G., \apj 352, (1990) 654-668
  
\bibitem{Scott:2008ns}
  Scott, P., Fairbairn, M. and Edsj\"o, J., \mnras 394, (2009) 82

\bibitem{LopesSilk2010Sc}
  Lopes, I. and Silk, J., Science 330, (2010) 462
  
\bibitem{2012ApJ...746L..12T}
  Turck-Chi{\`e}ze, S., Garc{\'{\i}}a, R., Lopes, I. \textit{et al.}, \apjl 746, (2012) L12

\bibitem{2012ApJ...752..129L}
  Lopes, I. and Silk, J., \apj 752, (2012) 129
  
\bibitem{CasanellasBrandao}
  Casanellas, J., Brand\~ao, I., and Lebreton, Y., \prd 91, (2015) 103535

\bibitem{art-RoxburghVorontsov2003A&A}
  Roxburgh, I. and Vorontsov, S., \aap 411, (2003) 215-220
  
\bibitem{BrandaoCunhaJCD}
  Brand\~ao, I., Cunha, M. and Christensen-Dalsgaard J., \mnras 438, (2014) 1751

\bibitem{SilvaAgDushera}
  Silva Aguirre, V., Basu, S., Brand\~ao, I. \textit{et al.}, \apj 769, (2013) 141

\bibitem{Rodriguez-Mdwarfs}
  Rodr\'iguez-L\'opez, C., Gizis, J. E., MacDonald, J., Amado, P. and Carosso, A., \mnras 446, (2015) 2613
  

\end{thebibliography}
\end{document}